\documentclass{IEEEtran4PSCC}
\usepackage{amsmath, amssymb, amsfonts} 
\usepackage{algorithm}
\usepackage[noend]{algpseudocode} 
\usepackage{booktabs}
\usepackage{graphicx}
\usepackage{subcaption}
\usepackage[font={small}]{caption}
\usepackage{placeins}
\usepackage{xcolor}
\usepackage{tabularx}

\algrenewcommand\algorithmicrequire{\textbf{Input:}}
\algrenewcommand\algorithmicensure{\textbf{Output:}}

\usepackage{booktabs}
\usepackage{multirow}
\usepackage{multicol}
\usepackage{comment}
\usepackage{hyperref}
\makeatletter
\let\old@ps@headings\ps@headings
\let\old@ps@IEEEtitlepagestyle\ps@IEEEtitlepagestyle
\def\psccfooter#1{%
    \def\ps@headings{%
        \old@ps@headings%
        \def\@oddfoot{\strut\hfill#1\hfill\strut}%
        \def\@evenfoot{\strut\hfill#1\hfill\strut}%
    }%
    \def\ps@IEEEtitlepagestyle{%
        \old@ps@IEEEtitlepagestyle%
        \def\@oddfoot{\strut\hfill#1\hfill\strut}%
        \def\@evenfoot{\strut\hfill#1\hfill\strut}%
    }%
    \ps@headings%
}
\makeatother

\psccfooter{%
        \parbox{\textwidth}{\hrulefill \\ \small{24th Power Systems Computation Conference} \hfill \begin{minipage}{0.2\textwidth}\centering \vspace*{4pt} \includegraphics[scale=0.06]{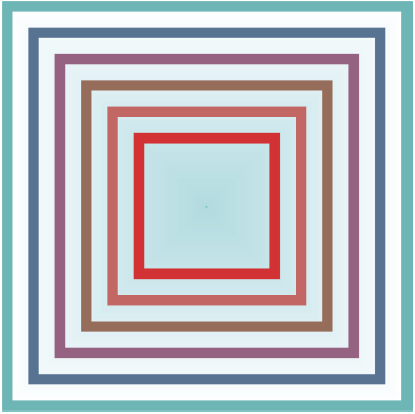}\\\small{PSCC 2026} \end{minipage} \hfill \small{Limassol, Cyprus --- June 8 -- June 12, 2026}}%
}

\begin{document}
%


\title{Bayesian Model-based Generation of Synthetic 
Unbalanced Distribution Networks Incorporating Reliability Indices}

\author{
\IEEEauthorblockN{
Henrique O. Caetano\IEEEauthorrefmark{1},
Rahul K. Gupta\IEEEauthorrefmark{2},
Cristhian G. da R. de Oliveira\IEEEauthorrefmark{1},
João B. A. London Jr\IEEEauthorrefmark{1},
Carlos Dias Maciel\IEEEauthorrefmark{3}
}
\IEEEauthorblockA{\IEEEauthorrefmark{1}Department of Electrical and Computing Engineering, University of São Paulo (EESC/USP), São Carlos, Brazil\\
henriquecaetano1@usp.br}
\IEEEauthorblockA{\IEEEauthorrefmark{2}School of Electrical Engineering and Computer Science, Washington State University, USA\\
rahul.k.gupta@wsu.edu}
\IEEEauthorblockA{\IEEEauthorrefmark{3}Faculty of Engineering and Science, São Paulo State University (UNESP), Guaratinguetá, Brazil\\
carlos.maciel@unesp.br}
}

\maketitle

\begin{abstract}
Real-world power distribution data are often inaccessible due to privacy and security concerns, highlighting the need for tools for generating realistic synthetic networks. Existing methods typically overlook critical reliability metrics such as the Customer Average Interruption Frequency Index (CAIFI) and the Customer Average Interruption Duration Index (CAIDI). Moreover, these methods often neglect phase consistency during the design stage, necessitating the use of a separate phase assignment algorithm. This work proposes a Bayesian Hierarchical Model (BHM) that generates phase-consistent unbalanced three-phase distribution systems, and incorporates reliability indices. The BHM learns the joint distribution of phase configuration, power demand, and reliability indices from a reference network, conditioning these attributes on topological features. We apply the proposed methodology to generate synthetic power distribution networks in Brazil, and validated it on known Brazilian networks. The results show that the BHM accurately reproduces the distributions of phase allocation, power demand, and reliability metrics on the training system. Furthermore, in out-of-sample validation on unseen data, the model generates phase-consistent networks and accurately predicts the reliability indices for the synthetic systems. The generated networks are also electrically feasible: three-phase power flows converge and voltages remain within typical operating limits, enabling studies of planning, reliability, and resilience.
\end{abstract}



\thanksto{\noindent Submitted to the 24th Power Systems Computation Conference (PSCC 2026).}

\section{Introduction}
The robust planning, analysis, and simulation of electric power distribution systems require detailed, granular network data, such as topology, component parameters, and power demand profiles \cite{li2022review}. However, such data is often inaccessible due to utility confidentiality policies 
and security concerns \cite{li2020building}. This data scarcity makes it difficult for researchers to evaluate scalability and to develop new methods and algorithms for modern distribution system planning and control.

To overcome this challenge, the generation of synthetic yet realistic distribution networks, has emerged as an essential alternative. Synthetic systems provide versatile datasets that mimic the characteristics of real-world networks without disclosing sensitive information, enabling a wide range of studies
\cite{li2020building, Caetano2024, Ali2023, Li2020, mateo2020building}. Crucially for power-system applications, synthetic networks must preserve electrical realism (e.g., plausible impedances, phase connectivity, and voltage behavior) to be useful for state estimation, optimal power flow, and reliability analysis~\cite{Caetano2025_powertech}.

Current methods for generating synthetic networks can be broadly categorized into two main approaches. The \textit{first} is a rule-based approach, which often uses publicly available georeferenced data from platforms like OpenStreetMap (OSM) combined with building data to define the network topology and estimate loads \cite{Ali2023, Li2020, Gupta2021}. While these methods can produce structurally realistic networks, they often provide a single deterministic output, making it difficult to quantify uncertainty. Furthermore, their reliance on fixed rules limits their flexibility and precludes the possibility of transfer learning, where knowledge from one system could be applied to another.

The \textit{second} approach utilizes statistical, data-driven tools to learn the characteristics of existing systems and then transfer this knowledge to generate new, unseen networks. This is particularly motivated by operators making distribution grid data available, such as in France \cite{enedis_cartographie}. This includes methods that model a distribution system as a graph to identify statistical patterns \cite{oneto2025large,tomaselli2024learning,Schweitzer2017} or use Bayesian models to probabilistically allocate network components \cite{Caetano2024}. A key challenge, however, is ensuring phase-consistency. While some methods have created unbalanced grids, they often rely on rule-based algorithms for phase assignment \cite{mateo2020building, Caetano2025_powertech}. Such an approach is not inherently data-driven and not suitable for probabilistic learning frameworks. 

Furthermore, the literature on synthetic networks has historically focused on operational aspects such as power demand and component impedance. A critical gap remains in the integration of reliability indices \cite{Fogliatto2022}. This work proposes developing reliability incorporated synthetic networks using statistical methods, where we associate the underlying distributions of failure events, specifically their frequency and duration, from real-world system data. 
By doing so, a generated network is endowed not only with operational parameters for simulation but also with realistic reliability characteristics, such as component failure rates. 
Providing these indices creates a self-contained test case with inherent failure probabilities derived from real-world operations, eliminating the need for users to rely on generic or arbitrary failure rates when conducting subsequent reliability and resilience studies.

This paper proposes a Bayesian Hierarchical Model (BHM) that addresses these gaps by generating synthetic, three-phase unbalanced distribution systems that are both phase-consistent by construction and incorporate reliability metrics. The key contributions of this work are as follows:
\begin{itemize}
    \item The \textbf{integration of reliability metrics} into the network generation process, allowing the model to learn the joint distribution 
    of key reliability indices such as Customer Average Interruption Frequency Index (CAIFI) and the Customer Average Interruption Duration Index (CAIDI).
    \item The development of a\textbf{ probabilistic framework that inherently models the phase configurations }of the network, ensuring sampled networks are consistent and feasible.
    \item A validation of the \textbf{model's transfer learning capability}. The BHM is trained on a reference system and subsequently applied to two unseen test systems with different power demand scales and reliability characteristics, demonstrating its ability to generate accurate, out-of-sample predictions.
    \item All generated samples \textbf{converge on three-phase power flow} and \textbf{adhere to standard voltage limit}s, supporting power engineering studies such as loss assessment, voltage regulation, and reliability planning.
\end{itemize}

The focus of this work is to generate parameters and metrics from the grid topology. While the topology is also difficult to obtain for low-voltage grids, it is out of scope for this work; readers can refer to other works that address it \cite{li2020building,Ali2023,banze2024open,pylovo_ref_2025}.
\section{Synthetic Network Generation Framework}
\begin{figure}[hbt!]
    \centering
    \includegraphics[width=\columnwidth]{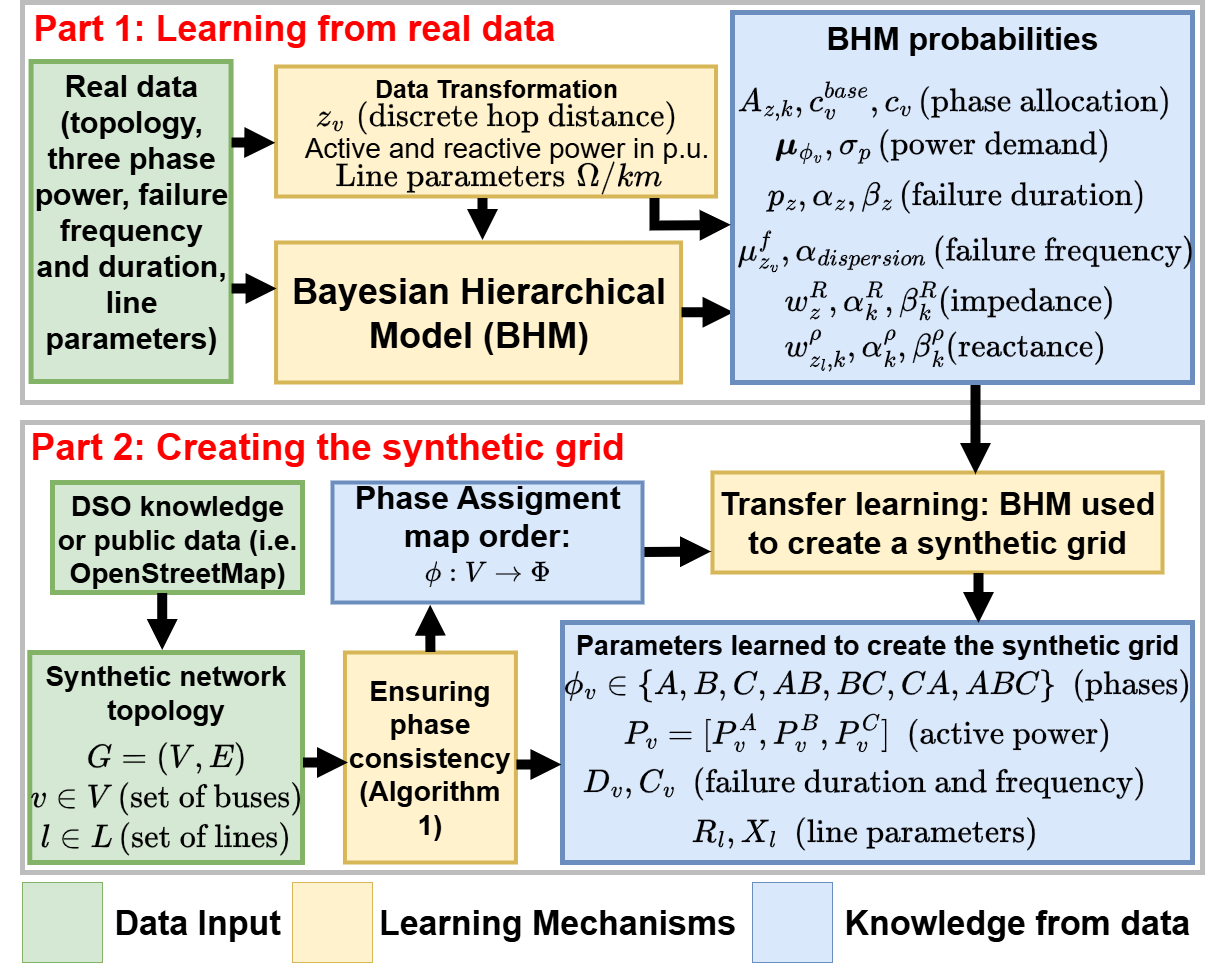}
    \caption{Flowchart of the proposed two-part methodology for generating synthetic distribution networks.}
    \label{fig:methodology_flowchart}
\end{figure}

The proposed synthetic network generation framework consists of two parts, as shown in Fig.~\ref{fig:methodology_flowchart}. The first part learns the statistical characteristics of a real-world distribution system using a BHM. Then, the second part applies this learned knowledge to generate a new synthetic grid, starting from a given network topology and ensuring physical constraints, such as phase consistency, are met. Regarding data input, the only requirement for the user is to provide the network topology, defined as the collection of buses, lines, and transformers. The framework treats each bus as a potential aggregation of customers, though users may also explicitly define no-load buses; in either case, the model automatically populates the electrical parameters and load profiles while strictly maintaining phase consistency across the provided structure.

At the core of this framework is the BHM, a statistical approach built on several key pillars \cite{mcglothlin2018bayesian}. \textit{First}, every unknown quantity, from model parameters to predictions, is treated as a random variable. Initial beliefs are specified as \textit{prior distributions}, which are updated via Bayesian inference to yield \textit{posterior distributions} once the model is fitted to observed data. \textit{Second}, this framework allows for the direct quantification of uncertainty by analyzing these posterior distributions. A key metric used is the Highest Density Interval (HDI), which represents the narrowest range containing a specified portion (e.g., 94\%) of the credible parameter values. \textit{Third}, the model is \textit{hierarchical}, meaning the distribution of one variable can be dependent on the parameters of another, which are themselves random variables. \textit{Fourth}, the BHM is able to efficiently sample from the learned posterior distributions. This allows for the rapid generation of a large ensemble of synthetic networks, where each complete network can be considered a single sample drawn from the model's learned understanding of the system.

To provide a concise overview of the generative framework, Table \ref{tab:model_summary} summarizes the probability distributions chosen to model each network parameter. The table details the specific predictors used for each component and outlines the statistical and physical reasoning that dictated these design choices.

\subsection{Model Components and Notation}
The BHM is designed to generate a comprehensive set of attributes for a synthetic network. The power distribution system is represented as a graph $G=(V,E)$, where $V$ is the set of buses (nodes) and $E$ is the set of lines (edges). Individual buses are indexed by $v \in \{1, \dots, |V|\}$, and individual lines are indexed by $l \in \{1, \dots, |E|\}$.

The primary topological predictor used throughout the model is a discretized measure of distance from the substation. For each bus $v$, we first calculate its continuous path distance in kilometers along the shortest path within the network graph from the substation to the bus. To capture potentially non-linear relationships between a component's location and its electrical or reliability characteristics, this continuous kilometer distance is then discretized into $Z$ distinct zones. We refer to the resulting categorical variable as the \textit{discrete hop distance zone}, denoted $z_v \in \mathcal{Z} = \{1, \dots, Z\}$. This approach adapts the concept of a topological hop count used in a previous work \cite{Caetano2024} to a measure based on physical distance. Similarly, the discrete hop distance zone for a line segment $l$, denoted $z_l$, is determined by the zone of its upstream node. It is important to highlight that researchers can use use other distance-based metrics within the framework by incorporating additional information about the grid. We use the shortest path to generate the synthetic grid using only the topology as input.

The BHM generates the following key characteristics for each component in the synthetic network:
\begin{itemize}
    \item \textbf{Phase Allocation ($\phi_v$):} The three-phase configuration for each bus $v$ (e.g., A, B, C, and their combinations).
\item \textbf{Active Power Demand ($\mathbf{P}_v$):} The per-phase power demand vector for each bus $v$, in kW.\footnote{The model also captures the reactive power demand per phase, which is formally defined later in Section~\ref{subsubsec:Hierarchical Power Model}.}
    \item \textbf{Interruption Duration (CAIDI, $D_{v}$):} The failure duration for each interruption at bus $v$ (hours).
    \item \textbf{Interruption Frequency (CAIFI, $C_v$):} The failure frequency for each bus $v$ (interruptions/year).
    \item \textbf{Line Parameters ($R_{l,1}, X_{l,1}$):} The positive-sequence resistance and reactance per kilometer for each line $l$ ($\Omega/\text{km}$.)
\end{itemize}
The BHM is structured as a generative process that models the aforementioned characteristics of a power distribution network, detailed in the following subsections.

\begin{table*}[htbp!]
    \centering
    \caption{Summary of Probabilistic Indicators and Distribution Choices in the Synthetic Network Framework}
    \label{tab:model_summary}
    \renewcommand{\arraystretch}{1.2} 
    \begin{tabularx}{\textwidth}{@{}l l l l X@{}}
        \toprule
        \textbf{Parameter} & \textbf{Symbol} & \textbf{Predictor} & \textbf{Distribution} & \textbf{Rationale \& Physical constraints} \\
        \midrule
        
        Phase Allocation & $\phi_v$ & Zone ($z_v$) & Dirichlet-Categorical & 
        Modeled as a probability vector that must sum to 1. Ensures non-negative probabilities for assigning phases (A, B, C, etc.) consistent with topology. \\
        
        Active Power & $\mathbf{P}_v$ & Phase ($\phi_v$) & Gamma & 
        Strictly positive distribution capable of capturing the right-skewed, heavy-tailed nature typical of electrical load data. \\
        
        Bus Deviation & $\Delta P_v$ & Phase ($\phi_v$) & Truncated Normal & 
        Allows individual bus loads to vary around the learned category mean while strictly enforcing non-negative power values. \\
        
        Interruption Duration & $D_v$ & Zone ($z_v$) & Hurdle-Weibull & 
        Captures the dual nature of reliability data: a binary "hurdle" for the probability of having zero interruptions, and a Weibull tail for the time-to-event nature of positive durations. \\
        
        Interruption Frequency & $C_v$ & Zone ($z_v$) & Negative Binomial & 
        Ideal for discrete count data (0, 1, 2...) that is overdispersed, where the variance of failure events exceeds the mean. \\
        
        Line Impedance & $R_{l}, X_{l}$ & Zone ($z_l$) & Gamma Mixture & 
        A mixture model accounts for heterogeneity in line types (e.g., overhead vs. underground). Gamma ensures strictly positive resistance values. \\
        
        \bottomrule
    \end{tabularx}
\end{table*}

\subsection{Phase Consistent Allocation Model}
\label{subsec:phase_allocation}
To create a realistic unbalanced distribution network, the phases assigned to each bus must be topologically consistent. This means the set of phases at any downstream bus must be a subset of the phases available at its upstream parent bus \cite{gupta_molzahn-phasebal_milp}. This section details a model that generates such physically plausible phase assignments in two main stages: 1) A hierarchical probabilistic model learns the likelihood of different phase configurations based on real data and 2) An algorithm uses the learned probabilities to assign phases, enforcing consistency by traversing the network's topological structure.
\subsubsection{Learning Phase Configuration Probabilities}
The first step is to build a statistical model that captures the probability of any given bus having a specific phase configuration. To model the likelihood of these states, we must learn a probability vector of length seven for each bus\footnote{As possible phases combinations are: \{A,B,C,AB,BC,CA,ABC\}}. The Dirichlet distribution is the natural choice for this task, as it generates probability vectors whose elements are non-negative and sum to one \cite{Caetano2024}.

We define a base probability vector, $\mathbf{c}_v^{\text{base}}$, for each bus $v$. This vector is drawn from a Dirichlet distribution whose concentration parameters, $\mathbf{a}_v$, are shared among buses located in the same hop distance zone $z_v$:
\begin{equation}
\mathbf{c}_v^{\text{base}} \mid  z_v \sim \texttt{Dirichlet}(\mathbf{A}_{z_v})
\end{equation}
\noindent A matrix of concentration parameters $\mathbf{A}$ of size $Z \times 7$ is defined, and is given a noninformative prior:
\begin{equation}
    A_{z,k} \sim \texttt{HalfNormal}(1) \quad \forall z \in \mathcal{Z}, k \in \{1,..,7\}
\end{equation}
\noindent to allow the model to learn from data.

The next section describes how these probabilities are used to ensure a consistent phase allocation.
\subsubsection{Topologically-Constrained Phase Allocation}
\label{subsubsec:phase_allocation}
With the learned probabilities in hand, the allocation algorithm assigns phases in a sequence that respects the network's physical structure. This process relies on identifying key branching points and propagating phase assignments, starting from the source feeder and going down the branches.

First, we define a structural hierarchy within the network graph $G=(V, E)$. We identify a set of ramification nodes, $\mathcal{R}$, which are defined as all nodes with a degree\footnote{the number of edges connected to that node} greater than two, plus the source node $v_s$:
\begin{equation}
    \mathcal{R} = \{v \in V \mid \text{deg}(v) > 2\} \cup \{v_s\}
\end{equation}
\noindent For each ramification node $r \in \mathcal{R} \setminus \{v_s\}$, we find the shortest path $\mathcal{P}_{r,s}$ to the source $v_s$. The parent of $r$, denoted $Pa(r)$, is defined as the ramification node on the path $\mathcal{P}_{r,s}$ that is closest to $r$. This creates a directed hierarchy rooted at the source, which defines the order of phase assignment.

The ramification nodes are assigned phases sequentially following the topological order, starting from the source $v_s$, which is typically assigned phase ABC. For each subsequent ramification node $r$, the phase $\phi(r)$ is sampled from a categorical distribution whose probabilities are derived from the learned base probabilities $\mathbf{c}_r^{\text{base}}$ but are constrained by the phase of its parent, $\phi(Pa(r))$.

To enforce this constraint, we define a transition function $\mathcal{T}(\phi_v)$ that returns the set of allowed downstream phases for a given phase $\phi_v$ that is already allocated for bus $v$. For example: If $\phi_v=AB$, then $\mathcal{T}(\phi_v)=\{A,B,AB\}$.

We then create a binary mask vector $\mathbf{m}_v$, where the $k$-th element is 1 if the $k$-th phase in $\Phi$ is in $\mathcal{T}(\phi(Pa(r)))$, and 0 otherwise. The final, constrained probability vector $\mathbf{c}_r$ is obtained by element-wise multiplication and renormalization:
\begin{equation}
    \mathbf{c}_r = \frac{\mathbf{c}_r^{\text{base}} \odot \mathbf{m}_r}{\sum_{k=1}^{7} (\mathbf{c}_r^{\text{base}})_k \cdot (\mathbf{m}_r)_k}
\end{equation}
where $\odot$ denotes the element-wise product. The phase for the ramification node is then sampled from this constrained distribution:
\begin{equation}
    \phi(r) \sim \texttt{Categorical}(\mathbf{c}_r)
\end{equation}

Once all ramification nodes have been assigned a phase, the remaining non-ramification nodes $v \in V \setminus \mathcal{R}$ are assigned phases deterministically. For each such node $v$, we find the shortest path $\mathcal{P}_{v,s}$ to the source. Its phase $\phi(v)$ is set to be identical to the phase of the closest ramification node $r^*$ found along this path towards the source.
\begin{equation}
    \phi(v) := \phi(r^*), \quad \text{where} \quad r^* = \arg\min_{r \in \mathcal{P}_{v,s} \cap \mathcal{R}} \texttt{dist}(v, r)
\end{equation}
This ensures that long branches originating from a ramification node carry the same phase configuration until the next major branch point. 

The complete process is summarized in Algorithm~\ref{alg:phase_allocation}.
\begin{algorithm}[!hbt]
\caption{Phase Consistent Allocation Algorithm}
\label{alg:phase_allocation}
\begin{algorithmic}[1]
\Require Network graph $G=(V,E)$, source node $v_s$, learned probability model $P(\cdot|z_v)$, transition constraints $\mathcal{T}$
\Ensure A phase assignment map $\phi: V \rightarrow \Phi$

\State Define ramification set $\mathcal{R} = \{v \in V \mid \text{deg}(v) > 2\} \cup \{v_s\}$
\State Determine parent relationships and topological order for all $r \in \mathcal{R}$ based on shortest paths to $v_s$
\State Initialize phase map $\phi$
\State Assign source phase: $\phi(v_s) \leftarrow \text{ABC}$

\For{each ramification node $r \in \mathcal{R} \setminus \{v_s\}$ in topological order}
    \State Get parent phase $\phi_p \leftarrow \phi(Pa(r))$
    \State Get base probabilities $\mathbf{c}_r^{\text{base}}$ from the learned model
    \State Define allowed phase set $\Phi' \leftarrow \mathcal{T}(\phi_p)$
    \State Construct mask $\mathbf{m}_r$ based on $\Phi'$
    \State Calculate final probabilities $\mathbf{c}_r \leftarrow \text{normalize}(\mathbf{c}_r^{\text{base}} \odot \mathbf{m}_r)$
    \State Sample phase: $\phi(r) \sim \texttt{Categorical}(\mathbf{c}_r)$
\EndFor

\For{each node $v \in V \setminus \mathcal{R}$}
    \State Find shortest path $\mathcal{P}_{v,s}$ from $v$ to $v_s$
    \State Find closest ramification on path to source: $r^* \leftarrow \arg\min_{r \in \mathcal{P}_{v,s} \cap \mathcal{R}} \text{dist}(v, r)$
    \State Assign phase deterministically: $\phi(v) \leftarrow \phi(r^*)$
\EndFor
\State \Return $\phi$
\end{algorithmic}
\end{algorithm}
\subsection{Hierarchical Power Model}
\label{subsubsec:Hierarchical Power Model}
The active power for each bus $v$, represented by the vector $\mathbf{P}_v = [P_{v,A}, P_{v,B}, P_{v,C}]$, is modeled conditionally on its assigned phase configuration, $\phi(v)$. We group the power profile of the buses in three categories: single-phase (A, B, C), dual-phase (AB, AC, BC), and three-phase (ABC).

To model the power consumption for each category, we use the Gamma distribution, as it is strictly positive and can the right-skewed, heavy-tailed nature often observed in load data \cite{Caetano2024}. Instead of learning independent parameters for each category, we use a hierarchical structure that assumes their statistical properties are related. The model implements this by having the specific shape ($\alpha_k$) and rate ($\beta_k$) parameters for each category's Gamma distribution drawn from a common parent Gamma distribution, which is itself defined by a single set of global hyperparameters ($\alpha_{hp}$, $\beta_{hp}$). This partial pooling of information allows the model to share the learning process across the categories, leading to more effective regularization and enabling robust learning even with few data samples for a particular category \cite{Caetano2025_powertech}. Building data is explicitly accounted for as a covariate in the prior distributions, as detailed in our previous work \cite{Caetano2024}.

For single-phase loads, we learn a single potential power value, $P^{\text{pot}}_{\text{mono}}$, which represents the characteristic power consumption of a typical single-phase bus. Its distribution is defined by:
\begin{align}
    \alpha_{\text{mono}} &\sim \texttt{Gamma}(\alpha_{hp}, \beta_{hp}) \\
    \beta_{\text{mono}} &\sim \texttt{Gamma}(\alpha_{hp}, \beta_{hp}) \\
    P^{\text{pot}}_{\text{mono}} &\sim \texttt{Gamma}(\alpha_{\text{mono}}, \beta_{\text{mono}})
\end{align}

This single scalar value is then used to construct the mean power vectors, $\boldsymbol{\mu}_k$, for the single-phase categories ($k \in \{A, B, C\}$):
\begin{align}
    \boldsymbol{\mu}_{A} &= [P^{\text{pot}}_{\text{mono}}, 0, 0] \\
    \boldsymbol{\mu}_{B} &= [0, P^{\text{pot}}_{\text{mono}}, 0] \\
    \boldsymbol{\mu}_{C} &= [0, 0, P^{\text{pot}}_{\text{mono}}]
\end{align}

Similarly, a total potential power, $P^{\text{pot}}_{\text{bi}}$, is learned for all dual-phase loads:
\begin{align}
    \alpha_{\text{bi}} &\sim \texttt{Gamma}(\alpha_{hp}, \beta_{hp}) \\
    \beta_{\text{bi}} &\sim \texttt{Gamma}(\alpha_{hp}, \beta_{hp}) \\
    P^{\text{pot}}_{\text{bi}} &\sim \texttt{Gamma}(\alpha_{\text{bi}}, \beta_{\text{bi}})
\end{align}

This total power must be distributed among the two active phases. We model this using a split factor, $\delta_{\text{bi}}$, drawn from a Beta distribution with a non-informative prior $\texttt{Beta}(2, 2)$. The mean power vectors for dual-phase categories ($k \in \{AB, AC, BC\}$) are constructed as follows:
\begin{align}
    \boldsymbol{\mu}_{AB} &= [P^{\text{pot}}_{\text{bi}} \cdot \delta_{\text{bi}}, \quad P^{\text{pot}}_{\text{bi}} \cdot (1-\delta_{\text{bi}}), \quad 0] \\
    \boldsymbol{\mu}_{AC} &= [P^{\text{pot}}_{\text{bi}} \cdot \delta_{\text{bi}}, \quad 0, \quad P^{\text{pot}}_{\text{bi}} \cdot (1-\delta_{\text{bi}})] \\
    \boldsymbol{\mu}_{BC} &= [0, \quad P^{\text{pot}}_{\text{bi}} \cdot \delta_{\text{bi}}, \quad P^{\text{pot}}_{\text{bi}} \cdot (1-\delta_{\text{bi}})]
\end{align}

For three-phase loads, the total potential power, $P^{\text{pot}}_{\text{tri}}$, is defined analogously:
\begin{align}
    \alpha_{\text{tri}} &\sim \texttt{Gamma}(\alpha_{hp}, \beta_{hp}) \\
    \beta_{\text{tri}} &\sim \texttt{Gamma}(\alpha_{hp}, \beta_{hp}) \\
    P^{\text{pot}}_{\text{tri}} &\sim \texttt{Gamma}(\alpha_{\text{tri}}, \beta_{\text{tri}})
\end{align}

To distribute this power among the three phases, we use a vector of split factors, $\boldsymbol{\delta}_{\text{tri}} = [\delta_1, \delta_2, \delta_3]$, which is modelled as a Dirichlet distribution with a non-informative prior: $\texttt{Dirichlet}([2, 2, 2])$. The mean power vector for the ABC category is:
\begin{equation}
    \boldsymbol{\mu}_{ABC} = P^{\text{pot}}_{\text{tri}} \cdot \boldsymbol{\delta}_{\text{tri}} = [P^{\text{pot}}_{\text{tri}} \cdot \delta_1, \quad P^{\text{pot}}_{\text{tri}} \cdot \delta_2, \quad P^{\text{pot}}_{\text{tri}} \cdot \delta_3]
\end{equation}

The previous step defines seven mean power vectors, $\{\boldsymbol{\mu}_k\}_{k \in \Phi}$. However, the power on each phase of a bus is allowed to vary from the mean. This variation is captured by a shared standard deviation parameter, $\sigma_p$:
\begin{equation}
    P_{v} \mid \phi_v \sim \texttt{TruncatedNormal}(\mu_{\phi_v}, \sigma_p^2, \text{lower}=0)
\end{equation}

Finally, the reactive power is calculated based on a constant power factor for the whole network ($PF$), modeled as three possibilities as shown in (\ref{eq:power_factor}), where $u$ is a random uniform distribution in the closed interval $[0,1]$, as previously used in the literature \cite{Caetano2024}.
\begin{equation}
\label{eq:power_factor}
\
    PF = 
\begin{cases}
    0.85,& \text{if } 0 < u \leq 0.1649\\
    0.90,& \text{if } 0.1649 < u \leq 0.27\\
    0.95,& \text{otherwise} \\
\end{cases}
\    
\end{equation}

\subsection{Interruption Duration Model}

To model the reliability of the synthetic network, we focus on generating customer-centric metrics directly at the bus level. While explicit building or customer density data could theoretically inform these metrics, our preliminary analysis of the training data indicated that such covariates yielded negligible improvements in predictive accuracy compared to topological factors. Specifically, the "hop distance" from the substation emerged as the dominant predictor, capturing the physical reality that downstream components are statistically more prone to interruptions due to cumulative exposure to failure risks. By relying on this topological metric rather than region-specific customer density data, the proposed model remains highly transferable to other regions where granular building data may be unavailable. Consequently, we model the interruption duration index (CAIDI) for each bus $v$ as a random variable conditioned on its hop distance zone, $z_v$. The same discussion also applied for the interruption frequency index (CAIFI).

CAIDI, which quantifies the duration of interruptions for each customer in the power system, is a continuous variable characterized by a significant number of zero-value observations (representing buses with no interruptions) and a continuous distribution of positive values (representing the duration of interruptions) \cite{Fogliatto2022}. To accurately capture this behavior, we employ a Hurdle-Weibull model \cite{deFreitasCosta2021}. This model is composed of two distinct parts: \textit{(a)} a binary hurdle component that models the probability of a failure occurring at all, and \textit{(b)} a continuous component that models the duration of the failure, given that it has occurred.

\paragraph{Hurdle Component (Zero vs. Positive CAIDI)}
First, we model the probability that a bus $v$ experiences any failure duration (i.e., has a CAIDI value greater than zero). Let $\lambda_v \in \{0, 1\}$ be a binary indicator variable where $\lambda_v=1$ if the CAIDI for bus $v$ is positive, and $\lambda_v=0$ otherwise. This is modeled as a Bernoulli process whose success probability, $p_v$, is dependent on the bus's hop distance zone, $z_v$. A separate probability, $p_z$, is learned for each zone from a Beta prior.
\begin{align}
    p_z &\sim \texttt{Beta}(1, 1) \quad \forall z \in \mathcal{Z} \\
    \lambda_v \mid z_v &\sim \texttt{Bernoulli}(p_{z_v})
\end{align}

\paragraph{Positive Value Component (Duration Magnitude)}
Second, for the subset of buses that have a CAIDI above 0 ($\lambda_v=1$), we model the interruption duration, $D_v$. As duration is a strictly positive, time-to-event-like quantity, we use the Weibull distribution \cite{Caetano2025_powertech}. The shape ($\alpha$) and scale ($\beta$) parameters of the Weibull distribution are modeled hierarchically, with a separate pair of parameters learned for each hop distance zone, $z_v$. These zone-specific parameters are given weakly informative Half-Normal priors as:
\begin{align}
    &\alpha_z \sim \texttt{HalfNormal}(1) \quad \forall z \in \mathcal{Z} \\
    &\beta_z \sim \texttt{HalfNormal}(1) \quad \forall z \in\mathcal{Z} \\
    & (D_v \mid z_{v}) \sim \texttt{Weibull}(\alpha_{z_v}, \beta_{z_v})
\end{align}
The complete statistical model for the CAIDI, $D_v$, for any given bus is thus a mixture of a Hurdle model - to model the presence of at least one interruption event - and a Weibull distribution - to model its duration - both conditioned on its topological location in the network.

\subsection{Interruption Frequency Model}
CAIFI, which quantifies the number of failure events for a bus over a given period (e.g., occurrences/year). As these are non-negative integers, and failure events often exhibit more variability \cite{Fogliatto2022}, we use the Negative Binomial distribution. This distribution is well-suited for overdispersed count data, where the variance can be larger than the mean \cite{linden2011using}.

The model is hierarchical, where the expected failure frequency for bus $v$, denoted $C_v$, is determined by its topological location in the network, specifically its hop distance zone, $z_v$. Here, $C_v \in \{0, 1, 2, \dots\}$ is the random variable for the observed failure count at bus $v$ and is defined as:
\begin{align}
    & \mu^f_z \sim \texttt{HalfNormal}(1) \quad \forall z \in \mathcal{Z} \\
    & \alpha_{dispersion} \sim \texttt{HalfNormal(1)} \quad \forall z \in \mathcal{Z} \\
    & C_v \mid z_{v} \sim \texttt{NegativeBinomial}(\mu = \mu^{f}_{z_v}, \alpha = \alpha_{\text{dispersion}})
\end{align}
\noindent where $\mu^{f}_{z_v}$ is the mean failure frequency for a bus $v$ located in zone $z_v$. Here, $\alpha_{\text{dispersion}}$ is the global overdispersion parameter that allows the variance of the failure counts to exceed the mean.

\subsection{Line Parameter Model}

The BHM is extended to generate the physical parameters of line segments, specifically the per-kilometer positive-sequence resistance ($R_{l,1}$) and reactance ($X_{l,1}$). The primary predictor for these parameters is the line's discrete hop distance zone, $z_l$, which is motivated by the physical property that lines closer to the feeder must accommodate higher currents and thus often have different impedance characteristics \cite{Wang2022}. Additionally, as detailed next, we define a mixture model for the line parameters to account for heterogeneity arising from different line models used, as well as differences between underground cables and overhead lines.

The Gamma distribution is selected as it is a flexible, positive-only distribution well-suited for physical parameters like resistance. Based on internal testing and previous experience \cite{Caetano2024}, the model assumes a mixture of $K=3$ components, representing distinct types of line impedances (e.g., low, medium, and high).

The model is structured with two main parts: a set of global component distributions and a set of zone-specific mixture weights. The parameters for the $K$ global Gamma components are defined hierarchically to enforce ordering and improve identifiability. The mean of each component, $\mu^R_k$, is modeled as an ordered sequence, and all components share a common coefficient of variation, $c_v^R$:
\begin{align}
    \mu^R_1 &\sim \texttt{HalfNormal}(1) \\
    \Delta^R_k &\sim \texttt{HalfNormal}(1) \quad \text{for } k \in \{2, 3\} \\
    \mu^R_k &= \mu^R_{k-1} + \Delta^R_k \quad \text{for } k \in \{2, 3\} \\
    c_v^R &\sim \texttt{HalfNormal}(0.5)
\end{align}
\noindent These parameters are then transformed into the shape ($\alpha^R_k$) and rate ($\beta^R_k$) parameters required by the Gamma distribution, as defined in (\ref{eq:gamma_transform}).
\begin{equation}
    \alpha^R_k = 1/(c_v^R)^2, \quad \beta^R_k = 1/((c_v^R)^2 \cdot \mu^R_k) \quad \forall k
    \label{eq:gamma_transform}
\end{equation}

The mixture weights, which determine the probability of a line belonging to each component in the mixture, are conditioned on the line's hop distance zone. A separate probability vector $\mathbf{w}^R_z$ is learned for each zone $z$ from a Dirichlet distribution.The final likelihood for the resistance of line segment $l$ is then a mixture model where the weights are selected based on the line's zone, $z_l$:
\begin{equation}
    R_{l,1} \mid z_l \sim \sum_{k=1}^{K} w^R_{z_l, k} \cdot \texttt{Gamma}(\alpha^R_k, \beta^R_k)
\end{equation}

An identical model structure is used to generate the X/R ratio, $\rho_l$. This ratio is a positive and typically low-valued number in power distribution systems, making the Gamma distribution a suitable choice for its components. Learning the ratio allows the model to capture typical conductor and line configurations. The likelihood for the ratio $\rho_l$ is:
\begin{equation}
    \rho_l \mid z_l \sim \sum_{k=1}^{K} w^\rho_{z_l, k} \cdot \texttt{Gamma}(\alpha^\rho_k, \beta^\rho_k)
\end{equation}
The final reactance value for the line segment is then deterministically calculated by:
\begin{equation}
    X_{l,1} = \rho_l \cdot R_{l,1}
\end{equation}

The framework generates topologically and load-unbalanced three-phase networks using positive-sequence parameters for the line segments. To enable detailed three-phase analysis, the full phase impedance matrix ($Z_{abc}$) is then deterministically constructed from these parameters using Carson's equations~\cite{Kersting2011}. For this calculation, we assume typical overhead conductor spacings of $d_{ab} = 0.6$~m, $d_{bc} = 0.6$~m, and $d_{ac} = 1.2$~m\footnote{These values are taken from the same data used to learn the BHM}. This approach was chosen because our attempts to learn the complete $Z_{abc}$ matrix directly struggled to generalize during transfer learning, likely due to uncertainties in identifying parameters at such a fine level of physical detail. Therefore, this approach ensures that the resulting systems remain electrically realistic for most distribution-level studies.

\section{Results and discussion}
We use the python library Pymc4 library to model the BHM. OpenDSS is used to simulate three-phase power flow. The Networkx library was used to calculate graph properties.


Below, we present the real-world datasets used for training and validation, which are used to build the BHM model, and then the model is applied to generate a synthetic network for two unseen regions, demonstrating the model's transfer learning capabilities.

\subsection{Data Description}
\label{subsec:data_description}
This study utilizes data from the Power Distribution Geographic Database (\textit{Base de Dados Geográfica da Distribuidora}, BDGD), a public dataset containing nationwide information on Brazil's electrical distribution systems \cite{bdgd_ref}. 
The database provides detailed information for modeling, including the topology, connection nodes, reliability indices and phase allocation for each customer. To ensure data consistency and mitigate risks from different operational practices, we consider systems within the state of São Paulo, operated by the same Distribution System Operator (DSO).

For the simulations below, we choose three distinct city-wide distribution systems of different scales and reliability conditions to provide a rigorous test of the BHM's generalization capabilities. All system were taken from metropolitan areas. For anonymity, these systems are referred to as Systems A, B, and C, defined below. 
\begin{itemize}
    \item \textit{System A} serves as the primary dataset for training the model. It is a large-scale network, comprising over 13,500 buses and serving more than 20,000 loads.
    \item \textit{System B} is used as a testing dataset and is characterized by, on average, lower failure durations and frequencies than System A. This medium-sized system consists of over 3,500 buses serving approximately 3,700 loads.
    \item \textit{System C} is the second testing dataset, characterized by, on average, higher failure durations and frequencies than System A. This system has over 2,500 buses and more than 3000 loads.
\end{itemize}

By testing the model on Systems B and C, we can evaluate its ability to not only interpolate within the range of the training data but also to extrapolate and make credible predictions for unseen networks with both better and worse reliability characteristics. \footnote{The framework ensures that the generated synthetic networks preserve the statistical characteristics of the reference system while maintaining full data confidentiality.}

Finally, for a better visualization of the phase consistency algorithm, synthetic networks are also generated for the IEEE 123-bus test case~\cite{ieeeResourcesx2013}.

\subsection{Model Validation}
\label{subsec:model_validation}

A phase consistency check was conducted for the Brazilian systems. Phase consistency is achieved in 100\% of the samples for all systems. This result is expected, as this constraint is explicitly enforced by the phase consistency algorithm (Algorithm 1) during the generation process; however, this check confirms its correct implementation.

It is also important to check if the model learns the probabilities for each phase combination. Table~\ref{tab:model_comparison_hdi} compares the empirical probabilities of each phase configuration from the dataset with the posterior estimates derived from the model. The posterior mean estimates show a strong correspondence to the real data proportions.  

\begin{table}[hbt!]
\centering
\caption{Validation of the phase allocation model on System A, comparing empirical probabilities with the BHM's posterior mean, 94\% HDI, and MAPE.}
\label{tab:model_comparison_hdi}
\begin{tabular*}{\columnwidth}{@{\extracolsep{\fill}} l ccccc}
\toprule
\textbf{Variable} & \textbf{Real} & \textbf{BHM (mean)} & \textbf{HDI 3\%} & \textbf{HDI 97\%} & \textbf{MAPE (\%)} \\
\midrule
$c^{base}_{v,A}$   & 0.142 & 0.142 & 0.136 & 0.147 & 0.00 \\
$c^{base}_{v,B}$  & 0.137 & 0.137 & 0.132 & 0.143 & 0.00 \\
$c^{base}_{v,C}$   & 0.131 & 0.131 & 0.126 & 0.137 & 0.00 \\
$c^{base}_{v,AB}$   & 0.187 & 0.186 & 0.180 & 0.193 & 0.53 \\
$c^{base}_{v,BC}$  & 0.143 & 0.143 & 0.138 & 0.148 & 0.00 \\
$c^{base}_{v,CA}$  & 0.223 & 0.221 & 0.214 & 0.227 & 0.90 \\
$c^{base}_{v,ABC}$   & 0.038 & 0.039 & 0.036 & 0.042 & 2.63 \\
\bottomrule
\end{tabular*}
\end{table}

Besides validating the phase consistency model for the Brazilian systems, it is also tested on the IEEE 123-bus benchmark system, to showcase the phase allocation visually. The validation focuses on the model's response to different prior beliefs about phase likelihoods, ensuring that the topological constraints are respected regardless of the underlying probability distribution. We consider four distinct scenarios by fixing the base probability vector, $\mathbf{c}_v^{\text{base}}$, to test the model's robustness under specific constraints:
\begin{itemize}
    \item[a)] An uninformative scenario where all seven phase configurations are equally likely: $\mathbf{c}_v^{\text{base}} = [1/7, \dots, 1/7]$.
    \item[b)] A scenario prohibiting single-phase loads, allowing only two-phase and three-phase configurations.
    \item[c)] A scenario prohibiting two-phase loads, allowing only single-phase and three-phase configurations.
    \item[d)] A scenario prohibiting three-phase (ABC) loads.
\end{itemize}

\begin{figure}[hbt!]
    \vspace{-0.5em}
    \centering
    \includegraphics[width=\linewidth]{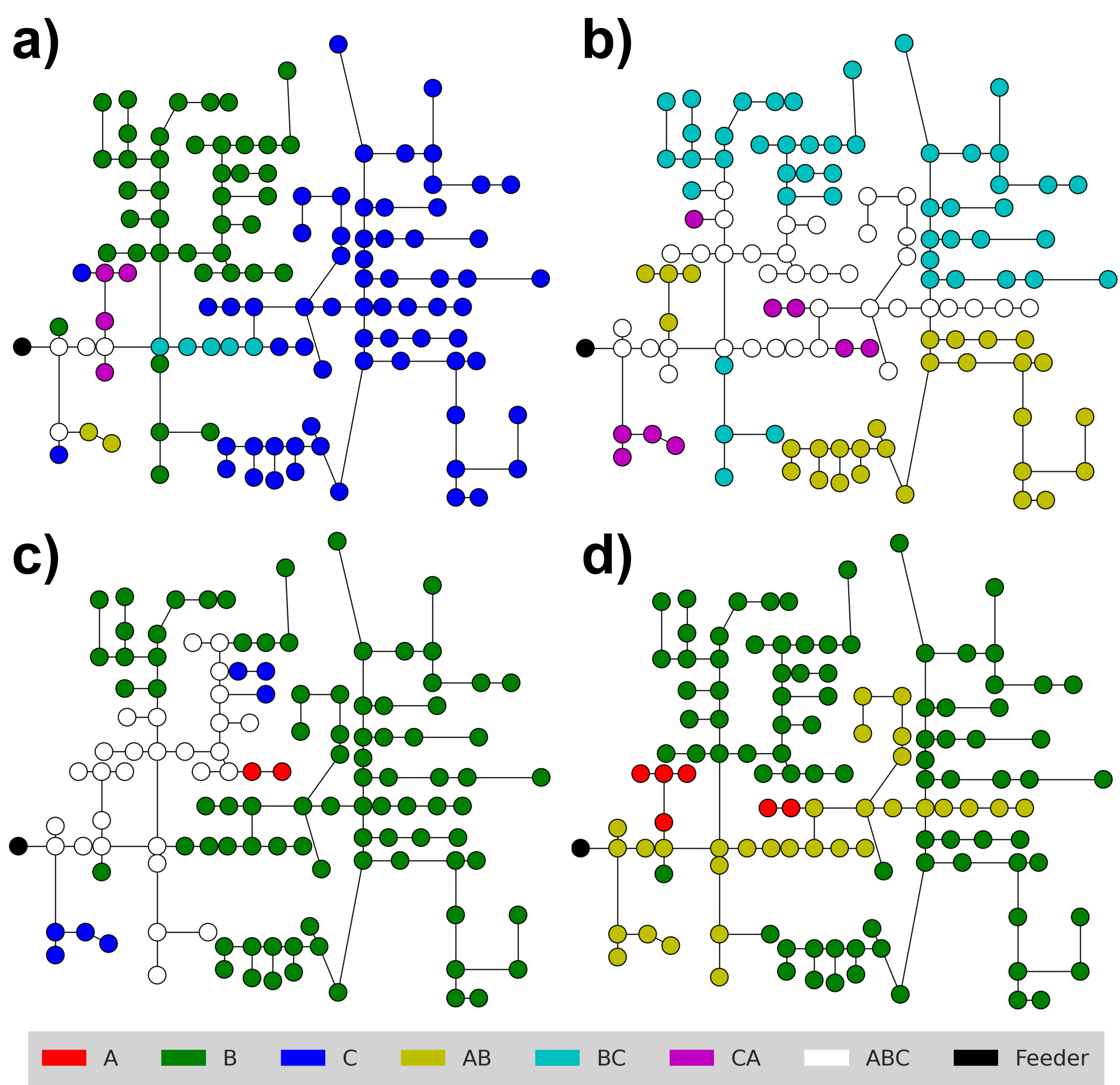}
    \caption{Phase allocation samples for the IEEE 123-bus system under four different prior probability scenarios. (a) Uninformative prior with all phases allowed. (b) Single-phase loads prohibited. (c) Two-phase loads prohibited. (d) Three-phase loads prohibited. In all cases, the resulting allocation is topologically consistent.}
    \label{fig:phase_allocation_ieee_123}
\end{figure}

Figure~\ref{fig:phase_allocation_ieee_123} presents a single stochastic realization from the BHM for each of these four scenarios. The results visually confirm that the model successfully generates a phase-consistent network in all cases. For instance, in scenario (c), no two-phase configurations (AB, BC, CA) appear, yet the single-phase branches correctly descend from valid three-phase parents. This demonstrates that the topological filtering mechanism correctly constrains the probabilistic assignments, ensuring the feasibility of the generated networks. The selection of phases is skewed, even with uniform probabilities, because the phase consistency algorithm limits or prohibits some phase combinations between adjacent nodes.

\begin{figure}[hbt!]
    \centering 
    \begin{subfigure}[b]{0.4\textwidth}
        \centering
        \includegraphics[width=\linewidth]{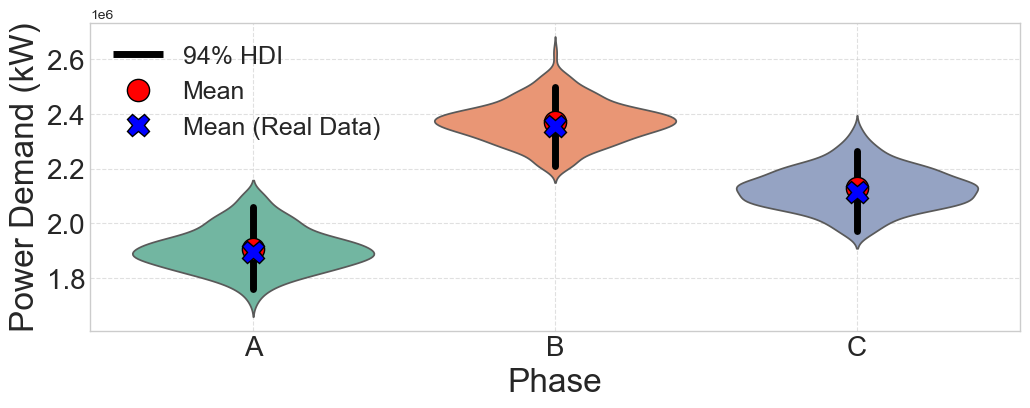}
        \caption{Validation on system A.}
        \label{fig:power_same}
    \end{subfigure}
    \hfill 
    \begin{subfigure}[b]{0.4\textwidth}
        \centering
        \includegraphics[width=\linewidth]{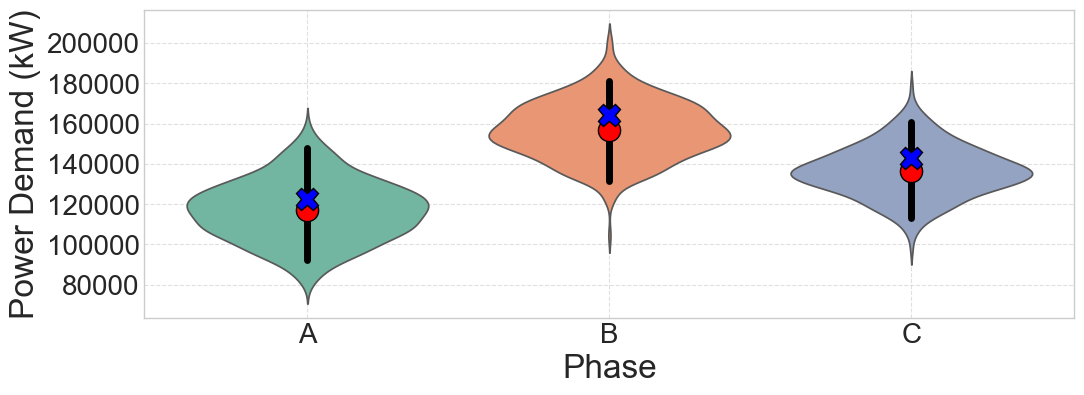}
        \caption{Transfer learning validation on system B.}
        \label{fig:power_tl_par12}
    \end{subfigure}
    \hfill 
    \begin{subfigure}[b]{0.4\textwidth}
        \centering
        \includegraphics[width=\linewidth]{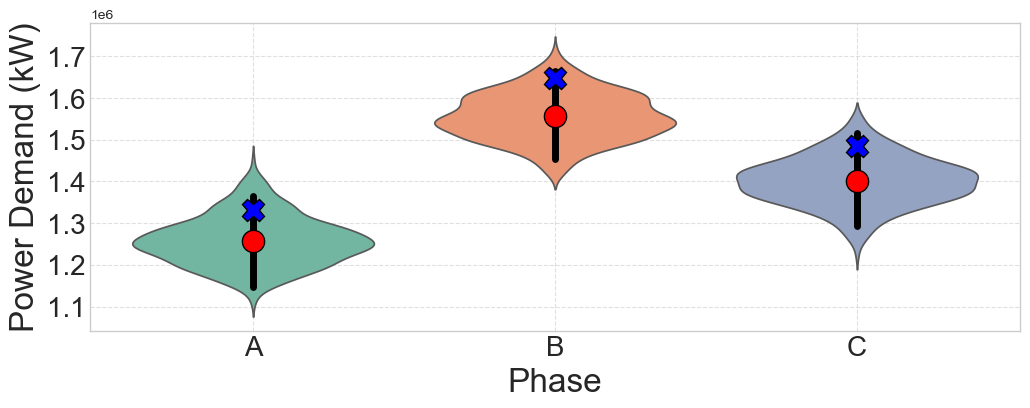}
        \caption{Transfer learning validation on system C.}
        \label{fig:power_tl_par14}
    \end{subfigure}
    \caption{Posterior predictive checks for the power demand. The figure compares the distribution (black line) and mean (blue cross) of real data  with the simulated distribution (shaded areas) and mean (red circle) from the BHM .}
    \label{fig:power_validation_combined}
\end{figure}

The validation for the power demand component of the BHM is presented in Figure~\ref{fig:power_validation_combined}. The analysis includes an in-sample check on the training data (System A) and out-of-sample transfer learning predictions for two unseen test systems (Systems B and C). 

For the in-sample case shown in Figure~\ref{fig:power_same}, the BHM samples accurately reproduces the characteristics of the observed data from System A. e also demonstrate the transfer learning capabilities by generating power profiles for two unseen systems, Systems B and C, as shown in Figures~\ref{fig:power_tl_par12} and \ref{fig:power_tl_par14}, respectively. The results indicate that the model generalizes effectively. This generalization is noteworthy because the target systems exhibit substantially different statistical properties from the training data. System B, for instance, operates at a much larger power scale, while its distribution shape differs in skewness from System A. 

\begin{figure}[htbp!]
    \centering 
    \begin{subfigure}[b]{0.45\textwidth}
        \centering
        \includegraphics[width=\linewidth]{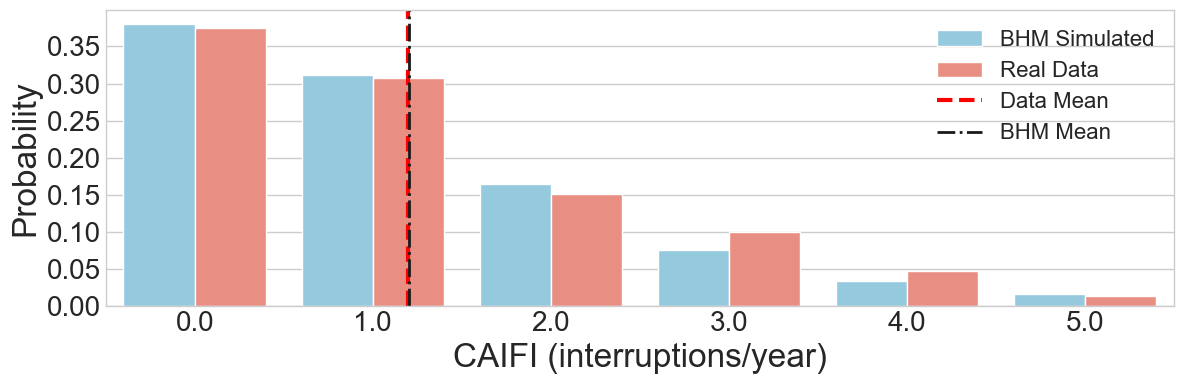}
        \caption{Validation on system A.}
        \label{fig:caifi_same}
    \end{subfigure}
    \hfill 
    \begin{subfigure}[b]{0.4\textwidth}
        \centering
        \includegraphics[width=\linewidth]{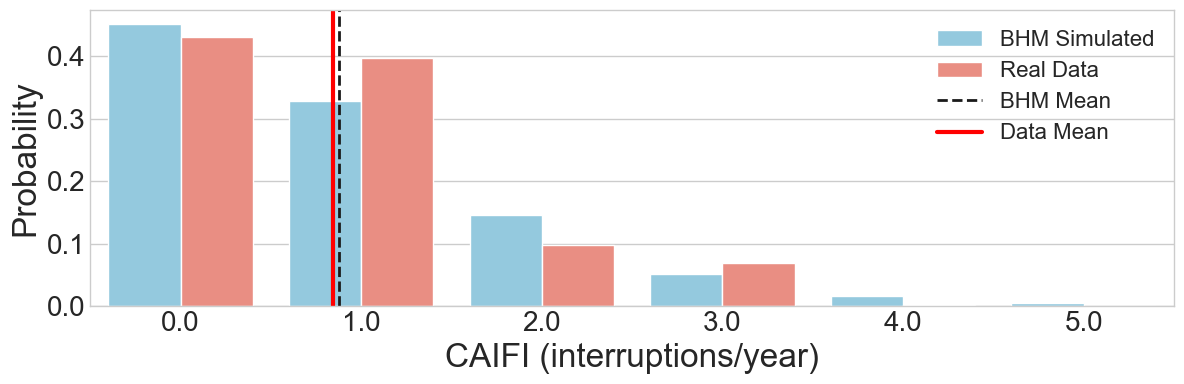}
        \caption{Transfer learning validation on system B.}
        \label{fig:caifi_tl_par12}
    \end{subfigure}
    \hfill 
    \begin{subfigure}[b]{0.4\textwidth}
        \centering
        \includegraphics[width=\linewidth]{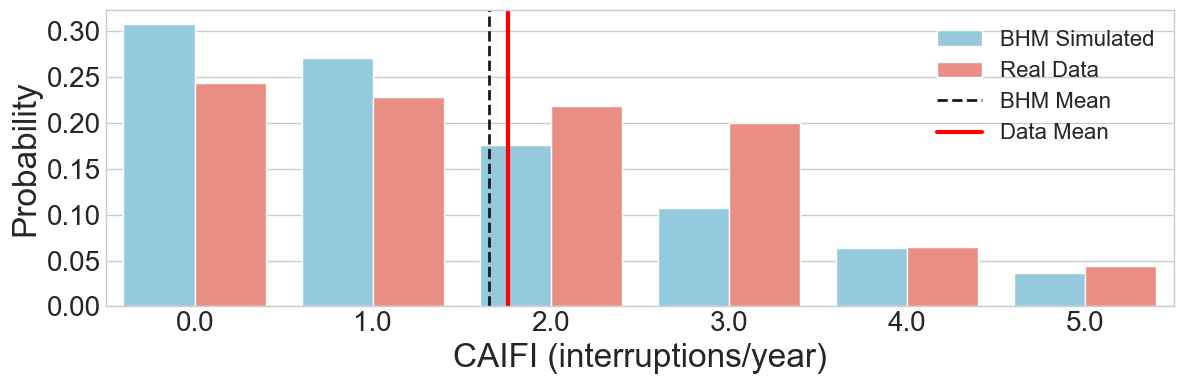}
        \caption{Transfer learning validation on system C.}
        \label{fig:caifi_tl_par14}
    \end{subfigure}
    \caption{Posterior predictive checks for CAIFI model. It compares the distribution of real data (black dashed line) with the simulated distribution from the BHM (blue shaded area).}
    \label{fig:caifi_validation_combined}
\end{figure}

The validation of CAIFI is shown in Figure~\ref{fig:caifi_validation_combined}.
Figure~\ref{fig:caifi_same} shows the in-sample validation on System A. As it can be seen, the distribution of the generated data by the BHM closely matches the observed data, and their respective means are close. This indicates a proper model fit to the training set.

The model's transfer learning capability was evaluated on System B (Figure~\ref{fig:caifi_tl_par12}), characterized by a lower average failure frequency than the training set, System A, and System C (Figure~\ref{fig:caifi_tl_par14}), which exhibits a higher average frequency. The results demonstrate that the model correctly captures these trends, predicting a lower mean CAIFI for System B and a higher mean CAIFI for System C. This confirms the model's ability to generalize and extrapolate based on the topological and electrical characteristics of the unseen networks.

While the model accurately predicts the mean CAIFI for System C, some discrepancies are visible in the proportions for individual frequency counts. This is a recognized challenge for data-driven models when predicting on unseen scenarios that lie outside the primary range of the training distribution. However, the central finding is that the model's prediction for the mean failure frequency is accurate for both out-of-sample systems. This confirms the model's ability to generate synthetic networks with realistic aggregate failure characteristics.

\begin{figure}[htbp!]
    \centering 
    \begin{subfigure}[b]{0.45\textwidth}
        \centering
        \includegraphics[width=\linewidth]{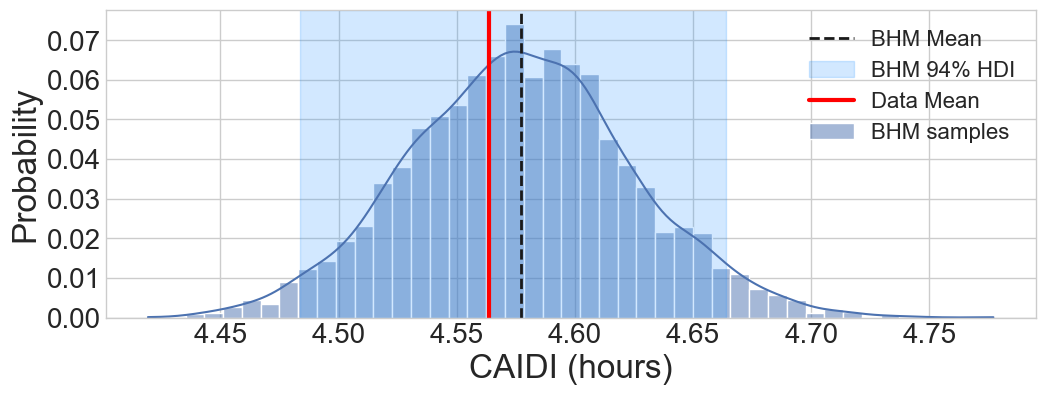}
        \caption{Validation on the training system.}
        \label{fig:caidi_same}
    \end{subfigure}
    \hfill 
    \begin{subfigure}[b]{0.45\textwidth}
        \centering
        \includegraphics[width=\linewidth]{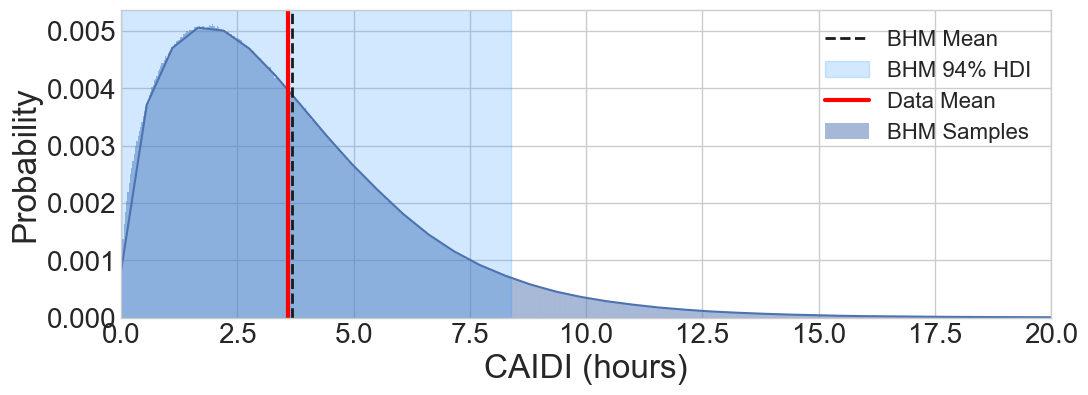}
        \caption{Transfer learning validation on System B.}
        \label{fig:caidi_tl_par12}
    \end{subfigure}
    \hfill 
    \begin{subfigure}[b]{0.45\textwidth}
        \centering
        \includegraphics[width=\linewidth]{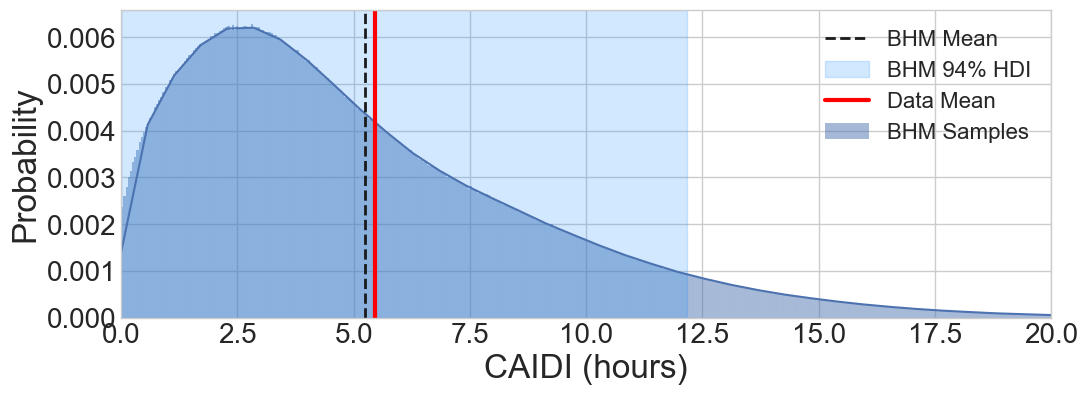}
        \caption{Transfer learning validation on System C.}
        \label{fig:caidi_tl_par14}
    \end{subfigure}
    \caption{Posterior predictive checks for CAIDI model. It compares the distribution of real data (black dashed line) with the simulated distribution from the BHM (blue shaded area).}
    \label{fig:caidi_validation_combined}
\end{figure}

A similar validation was performed for the interruption duration (CAIDI) model, with the results presented in Figure~\ref{fig:caidi_validation_combined}. The BHM correctly predicts a lower mean CAIDI for System B and a higher mean CAIDI for System C relative to the training data from System A. A key result is the high accuracy of the mean-level prediction across all three scenarios. In each case, the empirical mean of the real data is located centrally within the 94\% HDI of the model's posterior predictive distribution. This demonstrates that the BHM can generate synthetic data with accurate aggregate duration characteristics, even when extrapolating to systems with better or worse performance than the one it was trained on.

\begin{figure}[hbt!]
    \centering 
    \begin{subfigure}[b]{0.45\textwidth}
        \centering
        \includegraphics[width=\linewidth]{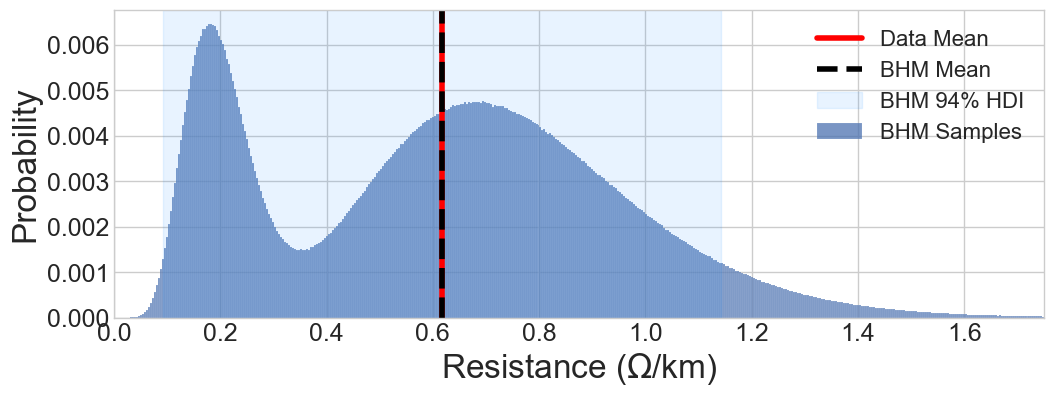}
        \caption{Resistance}
        \label{fig:line_param_resistance}
    \end{subfigure}
    \hfill 
    \begin{subfigure}[b]{0.45\textwidth}
        \centering
        \includegraphics[width=\linewidth]{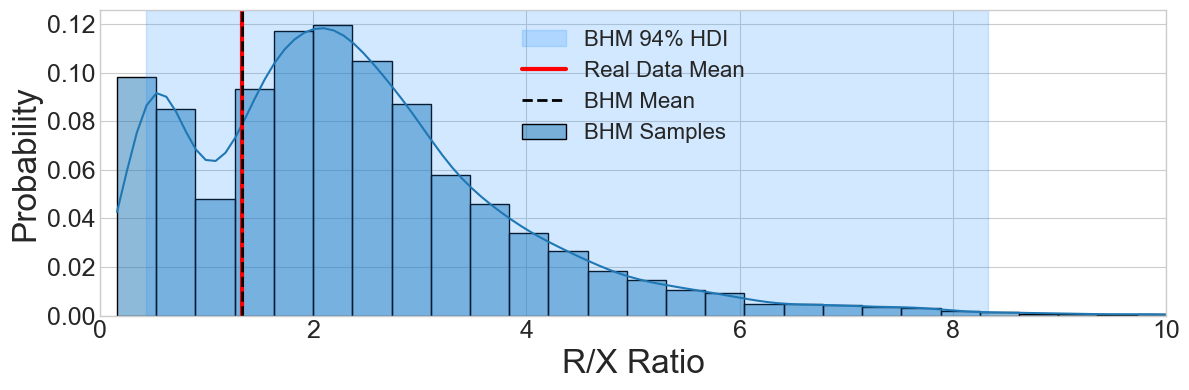}
        \caption{R/X ratio ($\rho_l$)}
        \label{fig:line_param_reactance}
    \end{subfigure}
    \hfill 
    \caption{Posterior predictive checks for the line parameters (resistance and R/X ratio). The figure compares the distribution of real data with the simulated distribution from the BHM.}
    \label{fig:line_param_validation}
        \vspace{-1em}
\end{figure}

The validation for the line parameter sub-model is presented in Figure~\ref{fig:line_param_validation}. The posterior predictive distributions for both the per-kilometer resistance ($R_{l,1}$) and the R/X ratio ($\rho_l$) show a close statistical alignment with the observed data. Both results highlight that the empirical distribution for resistance and reactance is bimodal, suggesting the presence of at least two underlying classes of conductors with different resistive properties. The hierarchical mixture model successfully identifies and learns this structure. 

\begin{figure}[hbt!]
    \vspace{-0.5em}
    \centering
    \includegraphics[width=\linewidth]{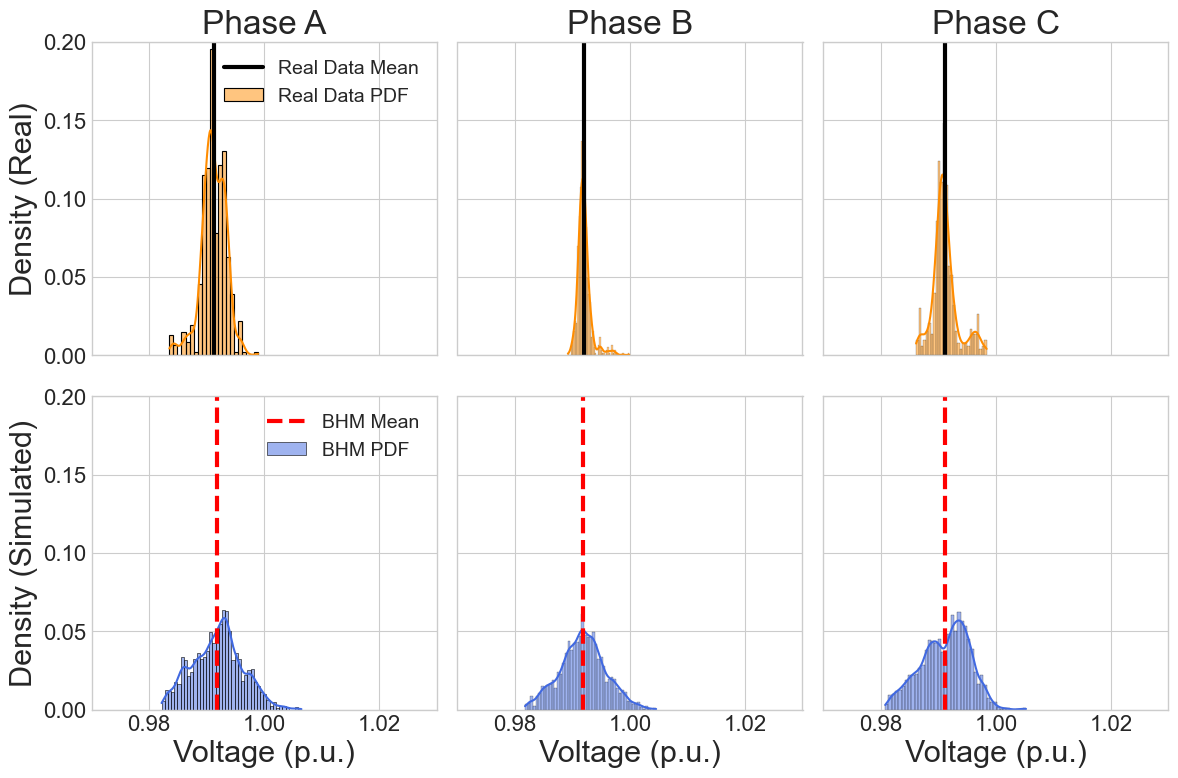}
    \caption{Comparison of measured and simulated voltage profiles for System B, separated by phase. The top row displays the voltage from the real data, while the bottom row shows the distribution of the generated samples from the BHM.}
    \label{fig:voltage_profile_opendss}
\end{figure}

It is important to highlight that the visualization strategy is determined by the data structure. Reliability indices (Fig.~\ref{fig:caidi_validation_combined}) and line parameters (Fig.~\ref{fig:line_param_validation}) represent static system snapshots, yielding a single aggregate mean for the real grid. These figures compare this single ground truth value against the BHM ensemble's posterior distribution to quantify model uncertainty. In contrast, Fig.~\ref{fig:voltage_profile_opendss} derives from a dynamic 8,760-hour simulation. Since the real system exhibits time-varying voltage magnitudes, a histogram comparison is necessary to validate the generated temporal statistics against the historical grid operation.

A power flow analysis was conducted on all samples, coupled with a voltage profile analysis in an 8760 study, considering the load profiles available in the original Brazilian dataset. resulting in a 100\% convergence rate across the three systems evaluated. Furthermore, all converged samples exhibited bus voltages within the standard operational limits of [0.9, 1.1]~p.u. Figure~\ref{fig:voltage_profile_opendss} provides a qualitative example of the voltage profile validation for System B, while also comparing it with the voltage profile from the real data. This result highlights that the synthetic grid generated has a voltage profile very similar to the original data. This is important since it allows for the generated grid to be used in voltage stability studies without loss of generality. It shows not only that the voltage level is adequate, but also that its distribution (i.e., slightly lower than 1 p.u.) follows that of the real grid.

%
\begin{figure*}[t]
    \centering
    \includegraphics[width=0.9\linewidth]{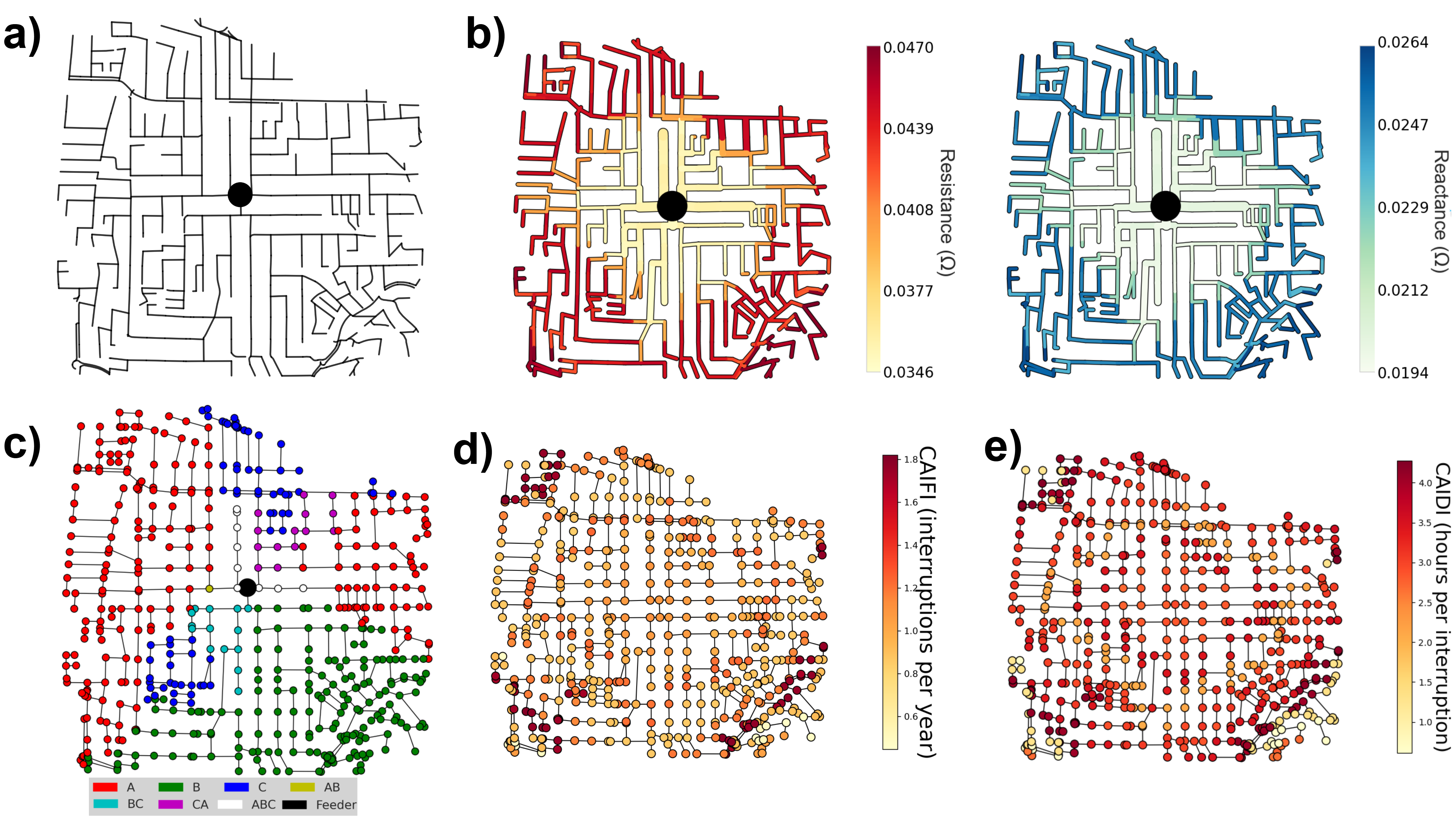}
    \caption{A single sample of a fully generated synthetic network for the central region of São Paulo, Brazil. The figure shows the (a) original OpenStreetMap region from which the network is generated (black dot represents the feeder); alongside the spatial distribution of (b) line parameters (the line width represent the number of phases), (c) phase allocation, (d) interruption frequency, and (e) interruption duration. The topology was derived from OpenStreetMap data \cite{Caetano2024}, and all electrical and reliability parameters were generated by the trained BHM.}
    \label{fig:from_scratch_sao_paulo}
\end{figure*}
\subsection{Generating New Synthetic Networks from the Ground Up}
In Section \ref{subsec:model_validation}, validation was conducted on Brazilian Grids, where the true data was available. In this section, to demonstrate the full capability of the BHM, the trained model was used to create completely new synthetic systems from the ground up. This process requires a pre-defined network topology which is derived using the OpenStreetMap data, approximating the road network as the power grid layout, a method applied in our previous work \cite{Caetano2024}. Then, we utilize the BHM to sample a full set of electrical and reliability parameters for every component in the system. To showcase the model's applicability to diverse real-world scenarios, synthetic networks were generated for the central region of São Paulo, Brazil.

Figure~\ref{fig:from_scratch_sao_paulo} presents a single stochastic realization of the generated network for the São Paulo case as a representative example. It is important to note that while the underlying probabilistic models generally capture increasing trends for parameters such as impedance and failure rates with respect to distance, the values shown here are the result of a single random draw. Consequently, local deviations may occur where downstream components exhibit lower values (e.g., lower $R$, $X$, or interruption indices) than their upstream counterparts due to the inherent variance of the sampled distributions.

For all generated networks, 100\% of the samples were phase-consistent, and a three-phase power flow converged for 100\% of the samples, indicating that the BHM generates networks with coherent and physically plausible parameters suitable for steady-state simulations. Since the voltage profile depends heavily on the load situation, an daily 8760 load profile case was conducted over all samples. The voltage profile statistics for all samples are summarized in Table~\ref{tab:voltage_stats}.  All buses operate within the standard [0.9, 1.1] p.u. limits.

\begin{table}[hbt!]
\centering
\caption{Voltage Profile Statistics (p.u.) for the synthetic network generated from the ground up for the central region of São Paulo, Brazil.}
\label{tab:voltage_stats}
\begin{tabular*}{\columnwidth}{@{\extracolsep{\fill}} l ccc}
\toprule
\textbf{Metric} & \textbf{Phase A} & \textbf{Phase B} & \textbf{Phase C} \\
\midrule
 Min     & 0.978 & 0.982 & 0.990 \\
 Mean    & 1.003 & 1.009 & 0.999 \\
 Max     & 1.019 & 1.023 & 1.012 \\
\bottomrule
\end{tabular*}
\end{table}

\subsection{Computational Time Analysis}
\label{}

The methodology is divided into two distinct stages: a one-time model training (learning) phase and a rapid synthetic network generation (sampling) phase. The generation of a new synthetic network is highly efficient, as detailed in Table~\ref{tab:comp_time}. The low computational cost of the sampling phase makes the proposed framework highly practical for a wide range of applications. The ability to generate thousands of plausible network instances in a short amount of time demonstrates the BHM's utility as a scalable and effective tool for power system analysis.
\begin{table}[hbt!]
\centering
\caption{Computational Time for BHM Processes. The synthetic network has over 10000 buses and lines.}
\label{tab:comp_time}
\begin{tabular*}{\columnwidth}{@{\extracolsep{\fill}} l r}
\toprule
\textbf{Process} & \textbf{Time} \\
\midrule
\textbf{BHM Training (One-time)} & \\
\quad All Models & 208 minutes \\
\midrule
\textbf{Synthetic Network Generation} & \textbf{(per sample)} \\
\quad Phase Allocation \& Power Demand & 200.0 ms \\
\quad Line Parameters (R1 \& X1) & 120.0 ms \\
\quad Interruption Duration (CAIDI) & 3.7 ms \\
\quad Interruption Frequency (CAIFI) & 3.3 ms \\
\bottomrule
\end{tabular*}
\end{table}
\section{Conclusion}
This work proposes a BHM for generating synthetic, three-phase unbalanced power distribution systems that are both phase-consistent by construction and include realistic reliability characteristics. We demonstrate the model's transferability by applying it to two distinct, unseen networks with different operational and reliability profiles. while the case studies utilize Brazilian data, the framework mitigates regional bias by relying on topological hop distance as a primary metric, allowing the model to naturally adapt to the structural characteristics of different grid topologies. Furthermore, the Bayesian formulation allows for the integration of region-specific input data, such as U.S. or European load profiles and grid codes, enabling operators to construct ensembles focused on their specific local environments. Finally, the generated cases are electrically feasible and immediately applicable to feeder studies (voltage regulation, loss analysis) and reliability planning. Researchers can use this tool to conduct worldwide studies, and grid operators of the modelled regions can also use the framework for individual studies and expansion planning.

Future research should focus on improving the model by integrating additional physical components, such as distributed energy resources and storage units. Furthermore, future iterations of the framework should move beyond purely local probabilistic decisions to incorporate global constraints, ensuring that line parameters reflect downstream loading, reliability metrics remain consistent across topology levels, and phase balancing is considered. The over-reliance of the proposed approach to distance-based metrics can also be further improved to account for important structural differences between physical and business contexts (rural versus urban feeders, for example). The line model can be further improved to explicitly account for underground cables and overhead lines. While the current reliability Bayesian model has been validated and demonstrated robust transfer learning capabilities, the reliability aspect could also be expanded to account for covariates other than hop distance, such as customer density and the effects of dynamic or localized weather events, enabling more thorough resilience studies. Investigating the scalability of this framework to larger, transmission-level systems presents an important opportunity for further exploration. Finally, regarding reproducibility, we intend to release a public Python package in a forthcoming tutorial paper that will include the data used for all Bayesian models and, crucially, allow users to input their own datasets to train custom region-specific models; in the meantime, the data used in this study is available upon request.

\section{Acknowledgements}
This work was partially fomented by São Paulo Research Foundation (FAPESP), grants 2021/12220-1, 2023/07634-7 and 2024/08485-8.

\bibliographystyle{IEEEtran}
\bibliography{bibliography.bib}
\end{document}